\documentclass[]{mn2e}
\usepackage{times}
\usepackage{epsfig}
\usepackage{amssymb}

\begin{document}
\title[Response of distance measures to the equation of state]
{Response of distance measures to the equation of state}
\author[Saini et al.]
{Tarun Deep Saini~$^1$, T.  Padmanabhan$^2$ and Sarah~Bridle$^1$ \\ 
$^1$ Institute of
Astronomy, University of Cambridge, Madingley Road, Cambridge CB3
0HA, United Kingdom\\
$^2$ Inter-University center for Astronomy and Astrophysics, Ganeshkhind, Pune 411 007, India 
} 
%
%\date{Accepted ???, Received ???; in original form \today}
%

%\keywords{cosmology: theory}
\maketitle

\begin{abstract}
We show that the distance measures (such as the luminosity and angular
diameter distances) are \emph{linear} function\emph{als} of the
equation of state function $w(z)$ of the dark energy to a fair degree
of accuracy in the regimes of interest.  That is, the distance
measures can be expressed as a sum of (i) a constant and (ii) an
integral of a weighting function multiplied by the equation of state
parameter $w(z)$.  The existence of such an accurate linear response
approximation has several important implications: (a) Fitting a
constant-$w$ model to the data drawn from an evolving model has a
simple interpretation as a weighted average of $w(z)$.  (b) Any
polynomial (or other expansion coefficients can also be expressed as
weighted sums of the true $w$.  (c) A replacement for the commonly
used heuristic equation for the effective $w$, as determined by the
CMB, can be \emph{derived} and the result is found to be quite close
to the heuristic expression commonly used.  (d) The reconstruction of
$w(z)$ by Huterer et al. (2002) can be expressed as a matrix
inversion.  In each case the limitations of the linear response
approximation are explored and found to be surprisingly small.

\end{abstract}

\section{Introduction}

Current cosmological data suggests that the expansion of the universe
is accelerating (e.g. Perlmutter et al. 1999; Riess et al. 1998;
Efstathiou et al. 2002; Lewis \& Bridle 2002; Melchiorri et
al. 2002). A number of models have been proposed to explain this fact,
the simplest of which is the cosmological constant. While a
cosmological constant is enough to explain the current data, its
constancy leads to a fine tuning problem (Sahni \& Starobinski 2000;
Peebles \& Ratra, 2002; Padmanabhan, 2002b).  A simple
phenomenological generalization of the cosmological constant is to
model the dark energy component that drives the acceleration as an
ideal fluid with an equation of state given by $P= w\rho$ in which the
equation of state parameter $w$ is allowed to vary with time. In this
parameterization the cosmological constant corresponds to $w=-1$,
while for other viable models $-0.6\lesssim w\lesssim -1$ at the
present epoch.  Such a parameterization indeed arises naturally in
several models, such as quintessence (Ratra \& Peebles 1988; Wetterich
1988), K-essence (Armendariz-Picon et al. 1999) and a tachyonic scalar
field (Gibbons 2002; Padmanabhan 2002a; Padmanabhan \& Roy Choudhury
2002; Bagla et al. 2003).  However the precise form of $w(z)$ is model
dependent.

In view of this there has been considerable interest in attempts to
summarize the current (and future) data in terms of a few
numbers. There are several ways of doing this, such as using a
polynomial approximation for $w(z)$ (e.g. Weller \& Albrecht, 2002) or
in terms of derivatives of the expansion factor (called Statefinders
by Sahni et al. 2002).  To be able to fit data using a polynomial form
we have to choose a low order polynomial since realistic data contains
only a finite amount of information.  On the other hand we need to use
enough polynomial orders so that the data is adequately described.
Saini et al. (2003) investigate in some detail the issue of how to go
about this and show that the current data and near future data seem to
require at most a low order polynomial. (This, of course, does not
imply that the true equation of state is also of a low order
polynomial form.)

These methods attempt to capture the effect of an unknown function
$w(z)$, which is equivalent to infinite number of parameters, in terms
of a finite (and often small) number of parameters.  In the limited
redshift range probed by supernova data the true variations in $w(z)$
might be such that a low order polynomial fit is a reasonable
approximation.  Alternatively one can expand the function $w(z)$ in
terms of a some set of basis functions which are complete in the given
redshift interval.  If the basis functions are chosen judiciously only
a small set of expansion coefficients will be required to describe the
function $w(z)$ within the limits of the experimental accuracy.  Owing
to the random noise in cosmological data any of these choices could
fit the data adequately in a maximum likelihood sense.  Since the
distance measures are integrals over nonlinear expressions involving
$w(z)$, any parameter determined by such a maximum likelihood analysis
represents some non-trivial average of the true $w(z)$.  Our main aim
in this paper is to show that the actual \emph{functional} relation of
the cosmological distance measures to $w(z)$ is not far from being
linear. This allows us to deduce an approximate relation between the
fitted coefficients and the true $w(z)$.  This allows one to obtain a
quantitative description of the averaging involved in reducing a
function $w(z)$ to a finite set of numbers.

The plan of this paper is as follows. In Section~2 we derive the
linear response approximation relating the coordinate distance and the
equation of state of the dark energy, and show that within a
reasonable range of parameters for the dark energy it works to better
than $\sim 2\%$.  In Section~3 we use this approximation to relate the
fitted coefficients of a polynomial form to the true equation of state
for the case of fitting to the luminosity distance, and show that they
are related through a weighted integral of the true equation of
state. We then generalize this result to the case of an arbitrary
functional form which is linear in the parameters.  In Section~4 we
derive a similar relationship for CMB where, in the simplest case,
only one measurement of the angular distance to the last scattering
surface is available. In Section~5 we show that the linear
approximation also enables a non-parametric estimation of $w(z)$. Our
conclusions are presented in Section~6.

\section{Response of geometry to the equation of state}
\label{sec:funct}

\begin{figure}
\vbox{\center{\centerline{
\mbox{\epsfig{file=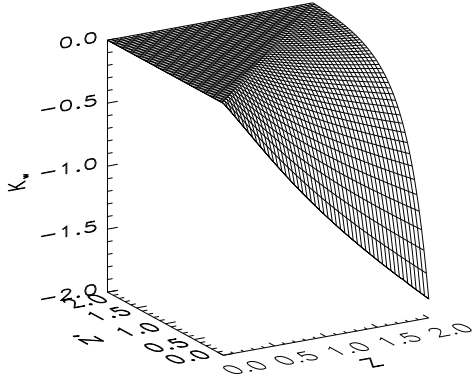,height=6cm}}
}}}
\caption{A surface plot of the kernel $K_w$ 
for supernovae distributed uniformly up to $z=2.0$. 
}
\label{fig:funct_surf}
\end{figure}

The luminosity and  angular diameter distances depend on the dark
energy through the coordinate distance $r[w,z]$ 
according to
the equations:
~$D_L(z) = (1+z) r[w,z]$ and $D_A(z) = r[w,z]/(1+z)$.  
The square brackets in
this notation
explicitly shows that at any redshift $z$ the coordinate distance
requires full knowledge of the function $w(z)$
In a flat universe the
coordinate distance $r[w,z]$ is given by
\begin{equation}
r[w, z] = (1+g)^{1/2} \,\int_1^{1+z} dx\, x^{-3/2}
[ g + Q[w,x] ]^{-1/2}
\end{equation}
\noindent
where $x=1+z$, $g = \Omega_{\rm m} / (1-\Omega_{\rm m})$,  
\begin{equation}
Q[w,z] = \exp
[3\int_1^{1+z} dx
\,w(x)/x ], 
\end{equation}
and we have set $c$ and $H_0$ equal to unity.  To explore the
behaviour of $r[w,z]$ when different equation of state functions
$w(z)$ are used, we need to understand the sensitivity of $r$ to
$w(z)$. This can be characterized by the
\emph{functional derivative} of $r$ with respect 
to $w(z)$. Since this is not a routine weapon in the arsenal of the
astronomer, we shall briefly introduce the concept before applying it.

In the case of a real function $f(x)$, the sensitivity of the function to the independent 
variable $x$ can be characterized by the derivative $df/dx = f'(x)$.  Broadly speaking,
a large value for $f'(a)$ indicates that $f$ is relatively more sensitive to the independent variable around the point
$x=a$; and a small value for $f'(a)$ will indicate relative insensitivity of $f$ to 
the independent variable around $x=a$. This follows directly from the definition of derivative of a 
function
\begin{equation}
f'(a) = \lim_{\epsilon\to 0} \frac{f(a+\epsilon) - f(a)}{\epsilon}\,\,.
\end{equation}
In the case of $f$ depending not on a single variable but on a \emph{function} $p(x)$ we need to 
study how $f$ changes if the function $p(x)$ is changed slightly around a point $x=b$.
This is best done by changing the function $p(x)$ to a new function $p_1(x)\equiv 
p(x) + \epsilon \delta_{\rm D}(x-b)$
which adds a ``spike'' at $x=b$ with a strength proportional to $\epsilon$.
We can now evaluate the value of $f$ for both $p(x)$ and $p_1(x)$.
The difference in the numerical values of $f$ in the limit of $\epsilon \to 0$ is a good measure
of the sensitivity of $f$ to the functional form of $p(x)$ around $x=b$. More 
formally, this functional derivative is defined by
\begin{equation}
\frac{\delta f[p,x]}{\delta p(b)} = \lim_{\epsilon\to 0} \frac{f[p+\epsilon \delta_{\rm D}(x-b),x] -f[p,x]}{\epsilon}\,\,.
\label{eq:deffuncder}
\end{equation}
Just as the ordinary derivative of a function depends on the location
at which it is evaluated, the functional derivative depends both on
the form of $p(x)$ around which the sensitivity is measured, $x=a$, as
well as the point $x=b$ at which the input function is perturbed.

For the purposes of this paper we need the response of the coordinate
distance $r[w,z]$ at a redshift $z$ to a change in the equation of
state at a different redshift, $z'$.  This can be computed around a
given fiducial $w(z) = w^{\rm fid}(z)$
from the functional derivative  
\begin{equation}
\frac{\delta r[w^{\rm fid},z]}{\delta w(z')} = 
\lim_{\epsilon \to 0} 
\frac{ r[w^{\rm fid} + \epsilon \delta_{\rm D}(z-z'), z]  
- r[w^{\rm fid},z]} {\epsilon}
\label{eq:derivative}
\end{equation}
defined exactly as in equation~(\ref{eq:deffuncder}) (also see Huterer
\& Turner, 2001 for a similar application of this idea)

For the rest of this section we switch from discussing the coordinate
distance to the luminosity distance and change the independent
variable from redshift $z$ to $x=1+z$.  Multiplying by $(1+z)$ we
obtain the response function for the luminosity distance $\delta
D_L[w^{\rm fid},x]/\delta w(x') \equiv K_w(x,x') $.  The subscript $w$
on the kernel denotes that it depends on $w^{\rm fid}$, the fiducial
equation of state around which the approximation holds. Evaluating the
expression in Eq~\ref{eq:derivative} we obtain
\begin{equation}
K_w(x,x')=
\left\{ 
\begin{array}{l}
-\frac{3x(1+g)^{1/2}}{2x'} 
\int_{x'}^{x} \frac{dy}{y^{3/2}} \frac{Q[w^{\rm fid}, y]}{\left (g+Q[w^{\rm fid}, y] \right )^{3/2}} \\
0 \quad \qquad (\mbox{for} \,\, x < x') 
\end{array}
\right. \,\,.
\end{equation}
The kernel $K_w$ is a function of two parameters, the redshift at
which $w(z)$ is perturbed and the redshift at which we consider the
change in the luminosity distance. A surface plot of this function is
shown in Fig~\ref{fig:funct_surf}. Since the effect of varying $w(z)$
at a redshift $z'$ is felt only at $z > z'$,the kernel is zero in half
the plane. For small $\delta w$ we can use this result to approximate
the calculation of the luminosity distance as
\begin{eqnarray}
\label{eq:approximation}
D_L[w^{\rm fid}+ \delta w, z] &\simeq& D_L[w^{\rm fid}, z] + \delta D_L \\
 &\equiv&D_L[w^{\rm fid}, z]+ \int_0^z K_w(z,z') \delta w(z') dz' \,\,.
\nonumber
\end{eqnarray}

\begin{figure}
\vbox{\center{\centerline{
\mbox{\epsfig{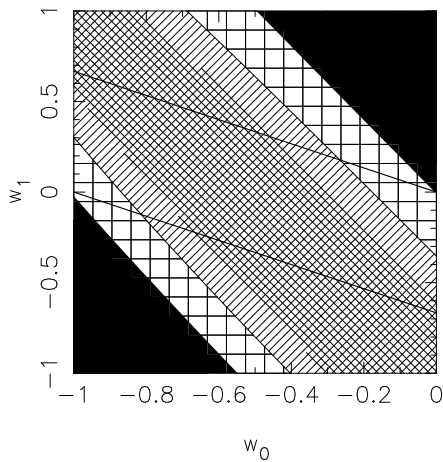}}
}}}
\caption{This Figure illustrates the accuracy of the linear
response approximation by plotting the relative difference in the
luminosity distance computed using the exact expression compared to
that from the approximate expression in Eq~7. This is calculated for
the special class of models in which the equation of state is a linear
function of redshift, $w(z)=w_0+w_1 z$.  The shaded regions give the
relative error $\delta d_l/d_l < 0.005, 0.01, 0.02, 0.05 $ from the
inner hatched region diagonally outwards. The two straight lines mark
the regions in which the $-1 < w(z) < 0$ in the range $0<z<1.5$.  }
\label{fig:contours}
\end{figure}

To be useful this linear response approximation should hold to a good
accuracy for a reasonable range of $w(z)$.  The SuperNova Acceleration
Probe (SNAP) survey is expected to observe about $2000$ Type 1a
supernovae (SNe), up to a redshift $z \sim1.7$, each year (Aldering et
al.\ 2002). A single supernova will measure the luminosity distance
with a relative error of $\sim 7\%$. If we bin the supernovae in
redshifts interval of $~0.02$, this will give a relative error in the
luminosity distance of about $\sim 1\%$.  Saini et al. (2002) show
that given this level of precision and given the present uncertainty
in the value of $\Omega_m$, the data seems to require at the most a
linear polynomial order in $w(z)$. To show how accurately
Eq~\ref{eq:approximation} gives the luminosity distance in this
restricted case, in Fig~\ref{fig:contours} we plot in the $w_0$-$w_1$
plane the percentage difference between the true luminosity distance
and the one computed through Eq~\ref{eq:approximation} at $z=1.5$. The
kernel $K$ was computed for $w^{\rm fid}(z) = -0.5$ in this
calculation.  Within the region $-1<w(z)< 0$ the approximation is
better than 2\%.  Therefore, the accuracy of the linear approximation
is quite acceptable with respect to the projected uncertainties in the
luminosity distance obtained from a SNAP class experiment. The
approximation works even better at smaller redshifts but less so at
higher redshifts, but the measurement errors are expected to be larger
here. Therefore, the conclusions drawn from this approximation are
expected to remain valid until an unprecedented accuracy is achieved
in the measurement of luminosity distance.

\section{Interpretation of the fitting parameters}

Since we have no fundamental understanding of the nature of dark
energy, it is necessary to assume some suitable, versatile, functional
form for $w(z)$ to fit the cosmological data. Since the luminosity
distance depends on the equation of state through an integral
relation, and the cosmological data is often fitted through maximum
likelihood method, we can always add a small fast varying term to any
well fitting $w(x)$, while still retaining a good fit
\footnote{ For example $w(x)$ and $w(x) +
\theta \sin (x/\theta)$, where $\theta \to 0$, would both fit the
data equally well though none of the derivatives agree. Although this
is a contrived example this point has been well illustrated more
realistically in Maor et al. (2000).}. Due to this integral dependence
any flexible parameterization eventually recovers only some integrated
property of the underlying equation of state. The simple
polynomial forms for $w(z)$ usually assumed for the purposes of
fitting the cosmological data are
often viewed as the first few terms in a series expansion of $w(z)$.
Since the behaviour of $w(z)$ is not necessarily polynomial like,
the recovered coefficients of the polynomial are not  related in a simple manner to
the true coefficients of the series expansion.
In this Section we use the linear
response approximation described above to compute the expectation
value for the coefficients of the polynomial approximating the $w(z)$,
and relate them to the  ``true'' input $w(z)$. This will show that such
fitting functions serve a useful purpose of measuring some integrated
properties of the true equation of state of the dark energy.

\subsection{Interpreting the fit with a constant $w$}

 The simplest   fit to the cosmological data is the
\emph{constant} $w$ model, $w^{\rm fit}(z)=w_0^{\rm fit}$. 
We are interested in relating this to the 
true
$w(z)=w^{\rm true} (z)$. 
For an arbitrary equation of state this relation is non-trivial and
non-linear but given the approximations considered in
\S~\ref{sec:funct}, we can construct the relationship
analytically. Let us first consider fitting the luminosity distance to
redshift. This requires a knowledge of the present day matter density
$\Omega_m$ as well and, for simplicity, we shall assume that this is
known. (When $\Omega_m$ is unknown, both $\Omega_m$ and $w$ become
biased; see Maor et al. 2001).  Our approach could be extended to the
case in which $\Omega_{\rm m}$ is not known, however this is beyond
the scope of the present paper.

In the maximum likelihood reconstruction
the quantity
\begin{eqnarray}
\chi^2 = \int_0^{z_{\rm max}} dz \,\, \left (\frac{D_L^{\rm fit}(z) - (D_L^{\rm true}(z) +n(z)) }{\sigma(z)}\right)^2
\label{eq:chisq}
\end{eqnarray}
is  minimized, 
where $n(z)$ is the noise in the measurement at redshift $z$ and
$\sigma(z)$ is the variance. (We have replaced the 
conventional
summation with an
integral.) 
We now approximate the fitting function and the luminosity 
distance by
\begin{eqnarray}
D^{\rm fit}_L(z)\hspace{-0.2cm} &\approxeq& \hspace{-0.3cm}
D_L[w^{\rm fid}, z] + \Delta w  \int_0^{z_{\rm max}} \! K_w(z,z')
dz'\\ 
D_L(z)\! \hspace{-0.2cm} &\approxeq& \hspace{-0.3cm}
 D_L[w^{\rm fid}, z] +\int_0^{z_{\rm max}}\! K_w(z,z'') \delta w(z'') dz'' 
\end{eqnarray}
where we have assumed a constant $w^{\rm fid}$ and  defined
$\Delta w = w^{\rm fit}_0 - w^{\rm fid}$ and $\delta w (z) = w^{\rm true}(z) -
w^{\rm fid}$. Minimizing $\chi^2$ and taking the expectation value we
obtain
\begin{eqnarray}
\Delta w = \int_0^{z_{\rm max}} \Phi_w(z') \delta w(z') dz'
\label{eq:linfit}
\end{eqnarray}
where
\begin{eqnarray}
\Phi_w(z) = \frac{\int \!\!\! \int
K_w(x,x')K_w(x,1+z)/\sigma^2(x)\,\, dxdx'}{\int\!\!\! \int\!\!\! \int 
K_w(x,x')K_w(x,x'')/\sigma^2(x) \,\,dxdx'dx''}
\label{eq:phidef}
\end{eqnarray}
where  all integrals are in the range $(0, z_{\rm max})$.
 The noise term has dropped 
out since its expectation value is zero.
Adding $w^{\rm fit}_0$ to both sides of Eq.~(\ref{eq:linfit}) we  obtain 
\begin{eqnarray}
w^{\rm fit}_0 = \int_0^{z_{\rm max}} 
\Phi_w(z') w^{\rm true}(z') dz' \,\,.
\label{fitisintofphi}
\end{eqnarray}
The fitted constant, $w^{\rm fit}$, is just a weighted average
of the true equation of state, with weighting function $\Phi_w$.
This weighting function is shown in Fig.~3 for the case of
supernovae distributed evenly from $z=0$ to $z=2.0$.
It decreases steadily with redshift, as might be expected from the fact that 
all the supernovae are affected by the equation of state at $z=0$, whereas the
equation of state at $z> 2.0$ affects no supernovae. 
This result shows that the value of $w^{\rm fit}_0$ obtained by
fitting to supernovae data remains invariant if the $w^{\rm true}(z')$
is changed in such a way that the weighted integral is unaffected.

If the linear response  approximation did not hold to a high accuracy then
Eq.~(\ref{fitisintofphi}) would lead to 
different results 
depending on the 
value for $w^{\rm fid}$ 
used 
to compute the weighting function
$\Phi_w$. Therefore a further test for the linear response
approximation is provided by the variation of the weighting function
evaluated for a different $w^{\rm fid}$. In Fig~3  we show the
weighting function computed for two different values of $w^{\rm fid}$,
namely $w^{\rm fid}=-1.0$ and $w^{\rm fid}=0$.  The figure shows that the
variation in the weighting function $\Phi_w$ is small.

To further quantify this difference we consider the change in 
$w^{\rm fit}_0$ with the change in the weighting function
\begin{eqnarray}
\Delta w^{\rm fit}_0 = \int_0^{z_{\rm max}} 
\Delta \Phi_w(z') w^{\rm true}(z') dz' \,\,.
\end{eqnarray}
Squaring both sides and using the Schwartz inequality we obtain the bound:
\begin{eqnarray}
(\Delta w^{\rm fit}_0)^2  \le 
\int_0^{z_{\rm max}} 
(\Delta\Phi_w)^2 dz \int_0^{z_{\rm max}} 
(w^{\rm true})^2 dz
\end{eqnarray}
Since we have assumed that $|w| < 1$  in the relevant range,   the second integral on the right hand side
has a maximum value $z_{\rm max}$. (We have computed the weighting
function for $z_{\rm max} = 2.0$.) With these numbers we can now
compute $(\Delta w^{\rm fit})^2$ by considering the difference of
$\Phi_w$ evaluated at $w_0= -1$ and $w_0=0$. We obtain $|\Delta w^{\rm
fit}| \sim 0.1$. This shows that the expected error in the
determination of $w$ through the linear response approximation is about $10\%$
if we linearize the luminosity distance function around an \emph{incorrect}
initial guess.  In practice, of course, one could do much better by first fitting
the cosmological data with a constant $w$ and then expanding around
that value so that the approximation works better than when it is
expanded about an arbitrary point.

\begin{figure}
\vbox{\center{\centerline{
\mbox{\epsfig{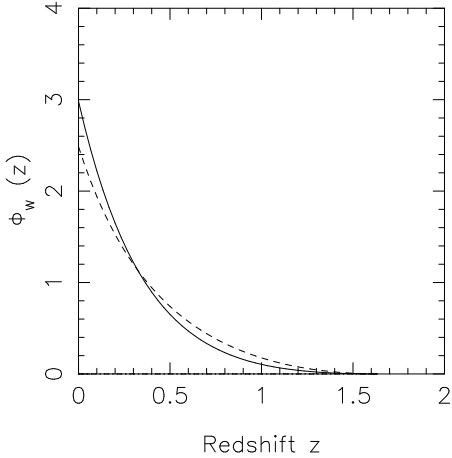}}
}}}
\caption{The weighting function for the constant $w$ estimator
for supernovae distributed uniformly up to $z=2.0$. The solid line shows the
weighting function computed with $w=-1$ and the dashed one for $w=0$. The small
difference between these plots shows that the linear approximation holds
to a good accuracy.
}
\label{fig:diffplot}
\end{figure}

\subsection{Generalization}
\label{sec:generalize}

We have derived above an expression for the expectation value of
$w^{\rm fit}$ for the case of a \emph{constant} $w$ fit.  The
corresponding expressions for the case of more complicated fitting
functions are, however, quite cumbersome.  We can simplify the
notation a little by considering discretized expressions.  Suppose the
luminosity distance is known at a large number of, uniformly distributed
redshifts ${z_i}$ where $i=1,N$. We consider a model $w(z)$ which is
given at redshifts ${z'_k}$ where $k=1,M$.

After having chosen a $w^{\rm fid}$, which for
simplicity is taken  to be a constant, we can define the vectors $
\mathbf {d} \equiv \{\delta D_L(z_i)\}$ and $ \mathbf {w} \equiv
\{\delta w(z'_i)\}$ and matrix $\mathcal{K} \equiv \{ K(z_i,z'_j) \delta z \} $, where
$\mathcal{K}$ is a $N \times M$ matrix and $\delta z$ denotes the
redshift interval between the bins.  With these definitions
Eq~(\ref{eq:approximation}) becomes \begin{equation} \mathbf{d} =
\mathcal{K}\mathbf{w}, \label{eq:matrixform} \end{equation} \noindent
where both ${\bf d}$ and ${\bf w}$ are small quantities.  Next we give
the equivalent discrete version of the weighting function obtained in
the previous subsection. We have $w^{\rm fit}(z) = w^{\rm fid} +
\Delta w$, where $\Delta w$ is a constant; therefore, using the
notation defined above we have ${\mathbf w}^{\rm fit} = {\mathbf
u}\Delta w$, where $\mathbf{u} = (1,1,\cdots,1)$, and similarly
$w^{\rm true} = w^{\rm fid} + \delta w(z)$, $\mathbf w^{\rm true} =
\{\delta w(z_i)
\}$. With these definitions 
Eq.~(\ref{eq:linfit}) and~(\ref{eq:phidef}) translate to 
\begin{eqnarray}
\Delta w= \frac{{{\mathbf u}^{\rm T} {\mathcal M} }} {{{\mathbf{u}^{\rm T}} 
{\mathcal M} {\mathbf u} }} \,\,{\mathbf w^{\rm true}}
\,\,;\quad {\mathcal M} = {\mathcal K}^{\rm T}_w \, {\mathcal K_w} \,\,.
\label{eq:disclinfit}
\end{eqnarray}

If instead we fit the data with a linear model $w^{\rm fit} = w^{\rm
fid} + \Delta w_0 + \Delta w_1 z$ then, following the same procedure
that led to Eq.~\ref{eq:linfit}, we find
\begin{eqnarray}
\Delta w_0 &=& \frac{\left [ {(\mathbf{z}^{\rm T}\mathcal{M}\mathbf{z})\mathbf{u}^{\rm T}\mathcal{M}  }  - 
(\mathbf{u}^{\rm T}\mathcal{M}\mathbf{z}) \mathbf{z}^{\rm
T}\mathcal{M} \right ] } {(\mathbf{u}^{\rm
T}\mathcal{M}\mathbf{u})(\mathbf{z}^{\rm T}\mathcal{M}\mathbf{z}) -
(\mathbf{u}^{\rm T}\mathcal{M}\mathbf{z})^2} \,\, \mathbf{w}^{\rm
true}\\
\Delta w_1 &=&  \frac{\left [{(\mathbf{u}^{\rm T}\mathcal{M}\mathbf{u})\mathbf{z}^{\rm T}\mathcal{M} 
- (\mathbf{u}^{\rm T}\mathcal{M}\mathbf{z})
\mathbf{u}^{\rm T}\mathcal{M}}\right ] } 
{ (\mathbf{u}^{\rm T}\mathcal{M}\mathbf{u})(\mathbf{z}^{\rm
T}\mathcal{M}\mathbf{z}) - (\mathbf{u}^{\rm
T}\mathcal{M}\mathbf{z})^2}
\,\,\mathbf {w}^{\rm true} \,\,,
\end{eqnarray}
where we have also defined the vector $\mathbf{z} = \{z_i\}$. This
generalizes the concept of the weighting function to the linear
case. It should be noted that the weighting function for $\Delta w_0$
is different  than in Eq.~(\ref{eq:disclinfit}),  since we are also
fitting for $w_1$.  However, as expected, these results agree for the special case
when $ \mathbf {w}^{\rm true} = \alpha \mathbf {u}$.

In the most general case we could approximate $w(z)$ as a linear
combination of an arbitrary set of functions $F_i$ 
(e.g. Gerke \& Efstathiou, 2002) as 
\begin{eqnarray}
w^{\rm fit}(z) = w^{\rm fid} + \sum_{i=1}^N  c_iF_i(z)
\end{eqnarray}
In this case the coefficients $c_i$s are related to the true equation 
of state as follows.
Defining the $N$ vectors ${\mathbf F}_i = \{F_i(z_k)\}$ we obtain
${\mathbf w}^{\rm fit} = \sum_{i=1}^N c_i{\mathbf F}_i$.
On performing a maximum likelihood analysis we find an expression for the 
coefficients $c_i$ as follows
\begin{eqnarray}
c_i = \sum_{i=1}^N a^{-1}_{ij} y_j
\end{eqnarray}
where we have defined the two matrices
\begin{eqnarray}
a_{ij} &=& {\mathbf F}_i^{\rm T} {\mathcal M} {\mathbf F}_j\\
y_i &=& {\mathbf F}_i^{\rm T}\mathcal{M}\mathbf{w}^{\rm true} \,\,.
\end{eqnarray}

\begin{figure}
\vbox{\center{\centerline{
\mbox{\epsfig{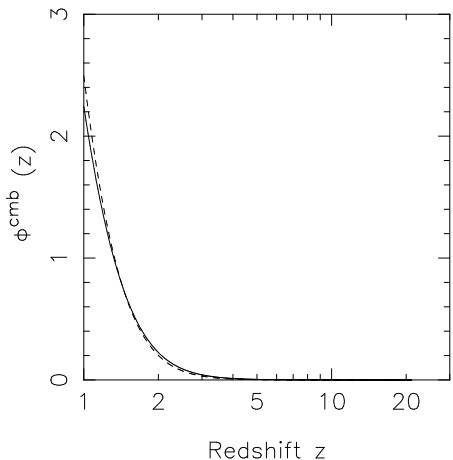}}
}}}
\caption{The weighting function which relates
the effective equation of state $w_{\rm eff}$ and the true equation of
state $w(z)$ as described in the text. The solid line shows the
weighting function computed from the linear response approximation
using $w^{\rm fid}=-1$ while the dotted one shows the weighting
function inferred from $\Omega_Q$. The two functions agree to a good
accuracy.  }
\label{fig:plot_cosmo}
\end{figure}

\def \ww{w}
\section{Effective $\ww$ seen by the CMB}

Current and near future supernova surveys will probe
redshifts only up to around $z=2$.
We can also compute the effective equation of
state probed  by the Cosmic Microwave Background (CMB) at $z_{\rm cmb}
\sim 1000$. In the simplest case the CMB gives us the angular distance to
the last scattering surface and therefore it does not give detailed
information on the behaviour of $w(z)$. Huey et al.\ (1999) show that
a 
$w_{\rm eff}$ that keeps the CMB temperature and the matter power
spectrum unaffected to an accuracy of $5$--$10\%$,  is given by
\begin{equation}
w_{\rm eff} = \frac{\int \Omega_Q(a)w(a)da}{\int
\Omega_Q(a) da}
\label{eq:heuristic}
\end{equation}
\noindent
where the integral extends up to the CMB redshift. This is
 in the same spirit as the weighting function corresponding to
fitting a 
constant constant equation of state described earlier.
Following a similar procedure 
to that in the previous section 
we obtain
\begin{eqnarray}
w_{\rm eff} &=& \int \Phi^{\rm cmb}(z) w(z) dz\\
\Phi^{\rm cmb}(z) &=& \frac{K'(1\!+\!z_{\rm cmb},1\!+\!z)}
{\int_0^{z_{\rm cmb}} K'(1\!+\!z_{\rm cmb},1\!+\!z) dz}
\end{eqnarray}
where $K'=K/(1+z)^2$ due to the difference between $D_L$ and $D_A$.

In Figure~\ref{fig:plot_cosmo} 
(solid line) 
we plot this weighting function as
calculated using the linear approximation by 
expanding around 
$w=-1$.
It has the same rough shape as the weighting function for supernovae
but this time extends to higher redshift. 
We see that despite the high redshift of the CMB the weighting function
falls off close to zero by redshift $z\sim 5$. This is due
to the fact that the dark energy term in the Friedman equations
is increasingly unimportant as we go to high redshifts.
For example it is well known that the cosmological constant is
dynamically unimportant at CMB redshifts.

\begin{figure}
\vbox{\center{\centerline{
\mbox{\epsfig{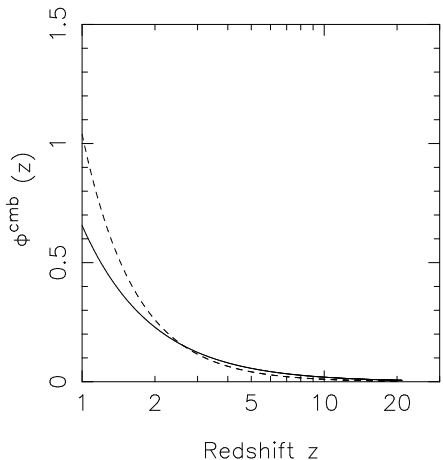}}
}}}
\caption{The weighting function for the CMB evaluated 
using $w^{\rm fid}=0$ (solid line).  The dotted line again shows the
weighting function inferred from $\Omega_Q$. The two functions do not
agree to a good accuracy, as discussed in the text.  }
\label{fig:plot_cosmo2}
\end{figure}

Overlaid on Fig.~\ref{fig:plot_cosmo} (dashed line) is the weighting function from
Eq.~\ref{eq:heuristic}, converting the integral from the scale factor
$a$ to redshift $z$. The good agreement shows that weighting with
$\Omega_Q$ works well for models close to the cosmological
constant. In Fig~\ref{fig:plot_cosmo2} we show the same plot but this
time expanding around $w^{\rm fid}=0$.  The two results do not agree
to a good accuracy. This shows that the weighting with $\Omega_Q$
should not work so well for models which have an average $w$ closer to
zero. These figures also show that the linear response approximation
works less well for the CMB since the weighting function changes by a
large amount when we expand around $w=-1$ and $w=0$. This means that
the weighting function would only apply for those models for which
$|w-w_{\rm eff}|$ is small.

\section{Non-parametric reconstruction of the equation of state parameter}
\label{sec:direct}

Although the fitting functions that we have  considered so far
provide useful information about the true equation of state of the 
dark energy, the linear approximation discussed above in principle allows 
a non-parametric reconstruction of the equation of state.

Consider a given SNe data set. To a first  approximation we can fit it with a
constant $w$, even though this may not be a good fit. 
We set the fiducial equation of state $w^{\rm fid}$ to this
best fit constant value in all that follows.
We next compute the
difference between the
SNe distances 
and the  $D_L$  computed for $w^{\rm fid}$ to obtain the residuals
$\delta D_L$.
We use the 
notation described in the first paragraph of
\S~\ref{sec:generalize}
so these 
are given by ${\mathbf d}^s$, where the superscript $s$ signifies that
these are obtained from measured supernovae distances. These are related to the
$w(z)$ in bins through Eq.~\ref{eq:matrixform}, where ${\mathbf
w } = \{w(z_k) - w_0\}$ with $w_0$ given by the constant $w$ fit.
Since
these are noisy estimates of distances we need to minimize the
$\chi^2$ function with respect to
$\mathbf w$ to obtain the Maximum Likelihood estimator for $\mathbf
w$;  that is, we minimize $\chi^2 = ({\mathbf d}^s - {\mathcal
K}{\mathbf w})^{T}({\mathbf d}^s - {\mathcal K}{\mathbf w})$
and obtain
\begin{eqnarray}
\mathbf w = ({\mathcal K}^{\rm T}{\mathcal K})^{-1}{\mathcal K}^{\rm T} {\mathbf d}^s\,\,,
\label{eq:linsolution}
\end{eqnarray}
where we have assumed the noise on $\mathbf{d}^s$ is constant with redshift,
although the result could be extended for the general case.

This equation also allows the trivial calculation of the Fisher matrix
${\mathcal F} = {\mathcal K}^{\rm T} {\mathcal K}$ corresponding to the
uncertainties on the reconstructed equation of state.  It was noted by
Huterer \& Starkman (2002) that the Fisher matrix was a surprisingly
weak function of the model parameters which matches with the result
obtained above and shows that it reflects on the validity of the 
linear response approximation.
 
Although Eq~\ref{eq:linsolution} gives a formal solution of the
problem, this estimation is, in general, very noisy. Huterer et
al. show that even for SNAP-like data there are only a few principal
components which are well determined by this method. It is clear that
this lack of resolution is largely due to the fact that no constraints
are imposed on the behaviour of $\mathbf w$. We discuss some of the
ways of rectifying this in a separate paper.

\section{Conclusions}

We have shown that in the relevant range of parameters expected for
the dark energy the luminosity distance is a linear functional to the
equation of state to a surprisingly high level of accuracy. This
approximation allows us to find the relationship between the usual
polynomial models for $w(z)$ and the true underlying equation of state
of the dark energy. Although the usual interpretation of these
polynomial approximation is in terms of a series expansion of $w(z)$,
we show that the coefficients of the polynomial approximation are
related to the true equation of state through an integral
relation. Only in the exact polynomial-like equation of state would
these approximations measure the real $w(z)$. We show that the fitting
of cosmological data with such forms is still useful since the
parameters of such models measure certain, well-defined, integrated
properties of the underlying equation of state. Finally, the
approximation allows a formal, non-parametric way to measure the
equation of state.

\subsection*{ACKNOWLEDGMENTS}

TDS and SLB acknowledges financial support from PPARC.  SLB also
thanks Selwyn College for support in the form of a Trevelyan Research
Fellowship.  We thank Irit Maor, Jochen Weller and Shiv Sethi for
helpful discussions.

\end{document}